\documentclass[aps, twocolumn,noshowpacs,preprintnumbers,amsmath,amssymb]{revtex4}

\usepackage{verbatim}
\usepackage{wasysym}

\usepackage{graphicx}
\usepackage{dcolumn}
\usepackage{bm}
\usepackage{wrapfig}

\begin{document}

\title{Oxygen and light sensitive field-effect transistors based on ZnO nanoparticles attached to individual double-wall carbon nanotubes}

\author{Alina Chanaewa}
\author{Beatriz H. Juarez}
\author{Horst Weller}
\author{Christian Klinke}
\email{klinke@chemie.uni-hamburg.de}
\affiliation{Institute of Physical Chemistry, University of Hamburg, 20146 Hamburg, Germany}

\begin{abstract} 

The attachment of semiconducting nanoparticles to carbon nanotubes is one of the most challenging subjects in nanotechnology. Successful high coverage attachment and control over the charge transfer mechanism and photo-current generation opens a wide field of new applications such as highly effective solar cells and fibre-enhanced polymers. In this work we study the charge transfer in individual double-wall carbon nanotubes highly covered with uniform ZnO nanoparticles. The synthetic colloidal procedure was chosen to avoid long-chained ligands at the nanoparticle-nanotube interface. The resulting composite material was used as conductive channel in a field effect transistor device and the electrical photo-response was analysed under various conditions. By means of the transfer characteristics we could elucidate the mechanism of charge transfer from non-covalently attached semiconducting nanoparticles to carbon nanotubes. The role of positive charges remaining on the nanoparticles is discussed in terms of a gating effect.

\end{abstract}

\maketitle

\section*{Introduction}

Zinc oxide (ZnO) is an amply investigated system with technologically relevant applications as piezoelectric transducers, optical waveguides, surface acoustic wave devices, varistors, phosphors, transparent conductive oxides, sensors, spin functional devices, UV-light emitters, and catalysts \cite{1,2,3}. With a direct band gap about 3.4 eV it is a promising material for photoelectric applications in the blue to UV range. Embedded transition metals can also turn ZnO into a diluted magnetic semiconductor with applications in spintronics \cite{4,5,6}. The growth of ZnO as nanoparticles (NPs) allows size, shape and surface control providing customized material for various applications \cite{7,8}. The synthesis in alcoholic solutions yields stable colloidal suspensions of ZnO nanoparticles with diameters in the range of a few nanometres. Additionally, continued heating of the nanoparticle dispersion can lead to nanorod formation by oriented attachment \cite{9}. 

Carbon nanotubes (CNTs) represent another system which exhibit extraordinary properties especially due to their one dimensional structure \cite{10}. The one dimensionality implies ballistic transport with conductance quantization, what makes CNTs ideal quantum wires. Their potential integration in transistors \cite{11,12,13}, conductive layers \cite{14}, field emitters \cite{15,16}, and mechanical components \cite{17,18,19} requires control over different electronic and structural properties. In the case of transistors, nanotubes serve as channels and therefore they need to be semiconducting. In contrast, for conductive layers metallic tubes are favoured. Unfortunately, the tube chirality which determines the band structure \cite{20} cannot be controlled by the CNT growth process at the present time but several post-synthetic separation protocols have been reported \cite{21,22,23,24,25}. However, structural features such as diameter and number of walls in a single tube are determined by the growth procedure; high-pressure carbon monoxide (HiPCO) process allows single-wall carbon nanotube (SWCNT) fabrication \cite{26}, laser ablation technique offers high degree of diameter control \cite{27}, and chemical vapour deposition (CVD) is a common method for large-scale production of multi-wall tubes \cite{28} and can be even used for secondary growth creating branched structures \cite{29}.

\begin{figure}[ht]
  \centering
  \includegraphics[width=0.43\textwidth]{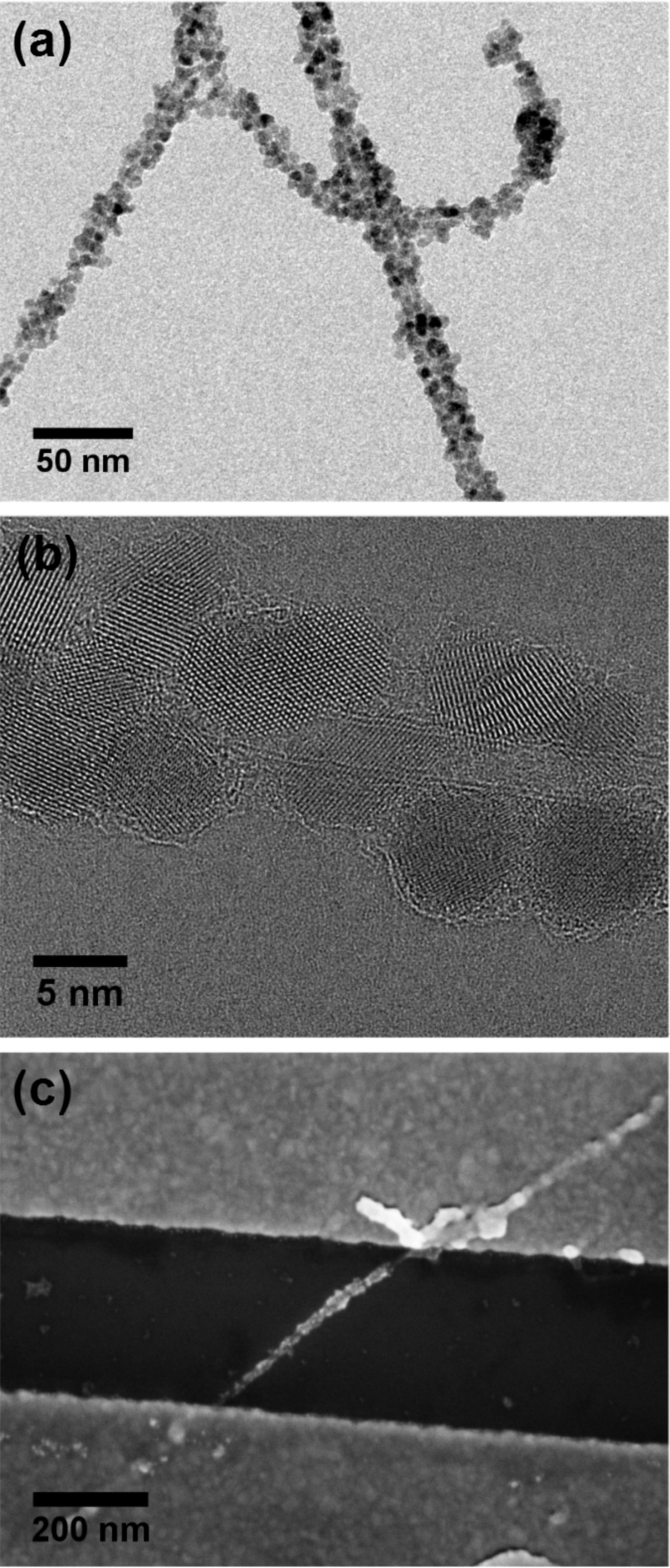}
  \caption{\textit{TEM images of ZnO nanoparticles obtained in the presence of DWCNTs: (a) an overview and (b) a high resolution micrograph. (c) SEM image of an individual ZnO-DWCNT contacted by gold leads forming source and drain in a FET device.}}
\end{figure}

The combination of nanoparticles, either metallic \cite{30,31,32} or semiconducting \cite{33,34,35}, with CNTs finds increasing interest due to the anticipation of synergetic effects. The attachment of metallic particles to CNTs allows plasmon coupling \cite{36} and opens new applications in photo-catalysis while the attachment of semiconducting ones shows clear benefits to the development of light harvesting assemblies \cite{37}. Several publications have demonstrated successful charge transfer between NPs and CNT \cite{38,39,40,41,42}. Usually this attachment involves covalent functionalization of the nanotubes \cite{43} or other harsh treatments \cite{39}, which degrades the optical and electrical properties of nanotubes \cite{13,44}. Newer approaches functionalize nanotubes non-covalently in order to attach various types of nanoparticles to the tubes \cite{45,46,47}. We recently reported a methodology to decorate non-functionalized CNTs by both semiconducting \cite{34,48,49} and metallic \cite{50} nanoparticles by non-covalent interactions. 

Vietmeyer et al. investigated ZnO-CNT composites, which were assembled via ZnO interaction with carboxylic groups of functionalized SWCNTs \cite{38}. A composite film was integrated in an electrochemical cell where it acted as absorber and charge collector. The resulting charge transfer paths were monitored via spectroscopic studies, which allow valid conclusions only for the cooperative behaviour of the composite layer opening space for more detailed research. Liu et al. reported an increase in photoresponsivity of an individual SWCNT decorated with ZnO nanoparticles by using dodecanoic acid as binder molecule \cite{45}.

In this work we have fabricated a well-defined ZnO-CNT field-effect transistors (FETs) using individual non-functionalized CNTs, avoiding long chain ligands at the interfaces. The composite material is obtained in colloidal solution, what allows the generation of high quality and monodisperse nanoparticles in milligrams to grams of product and high coverage of CNTs. The small radii of single-wall tubes favour bundling in organic solvents \cite{51} preventing ZnO attachment to individual nanotubes. For this reason, we chose double-wall carbon nanotubes (DWCNTs) since they can be dispersed individually in the organic solvent which is necessary for the colloidal synthesis. The obtained composites were deposited on the Si/SiO$_{2}$ wafer and metallic leads were defined using electron beam lithography. Photoelectrical investigations were performed under ambient conditions as well as in vacuum. To the best of the author knowledge, this is the first transfer characteristic study on ZnO nanoparticle covered DWCNT transistors.

\section*{Methods}

Synthesis of the composites: In order to debundle the DWNTs a dispersion of 0.2 mg DWCNTs (as received, Nanocyl SA, Belgium) was sonicated in 1 mL 2-phenyl ethanol for 30 min prior to injecting this suspension into a solution containing 270 mg zinc acetate dihydrate in 9 mL 2-phenyl ethanol. After reaching 60 $^{\circ}$C under stirring 6.5 mL of a potassium hydroxide solution (1.1 g KOH in 50 mL 2-phenyl ethanol) are injected. The reaction proceeds under ambient conditions at 60 $^{\circ}$C. After 120 min of synthesis the obtained composite was washed several times with methanol.

Device preparation: Several devices were prepared by dispersing the DWCNTs in 1,2-dichloroethane for blank experiments or ZnO-DWCNTs in methanol and drop-casting the material on a Si/SiO$_{2}$ wafer (100 nm SiO$_{2}$). Subsequently, a resist was deposited on the wafer and windows for source and drain electrodes were opened via electron beam lithography. To ensure an intimate contact between the CNT and the electrodes the resist-free surface was washed with water which selectively removes the nanoparticles at the contacts since ZnO-NPs are soluble in water. After thermal evaporation of 1 nm titanium and 20 nm gold, the remaining resist was lifted off by acetone. This process defines the electrodes.

The devices were contacted in a VFTTP4 probestation by LakeShore and illuminated with a 400 W Xenon lamp. The measurements were performed with a 4200-SCS semiconductor characterization system from Keithley Instruments. For the spectroscopically resolved photo-conductivity measurements a Photon Technology International monochromator (grating 600 L/mm) was used.

\begin{figure}[ht]
  \centering
  \includegraphics[width=0.45\textwidth]{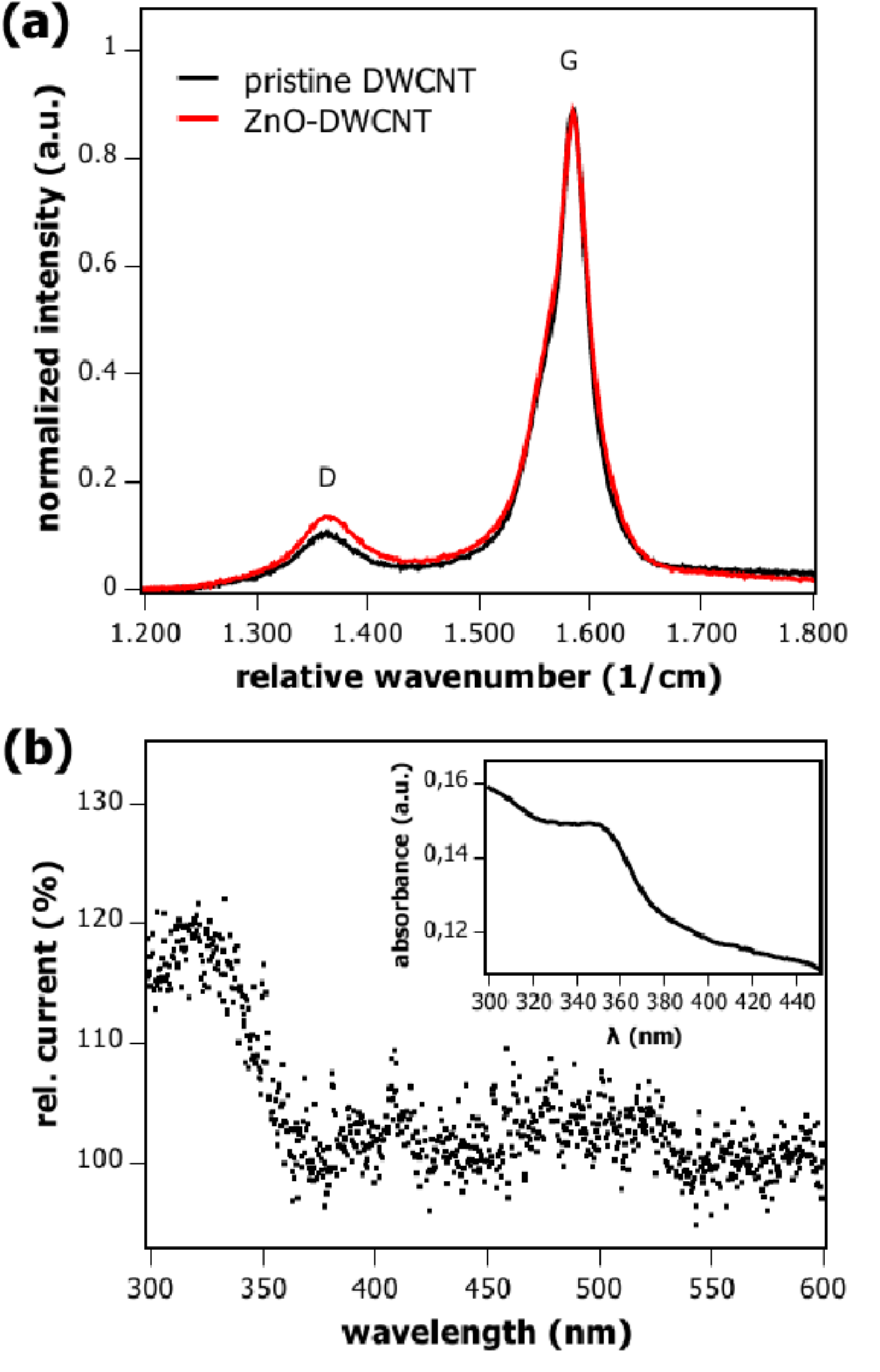}
  \caption{\textit{(a) Raman spectra of pristine DWCNTs and ZnO-DWCNTs composites shown in Figure 1a,b. Both curves are scaled to the G peak. (b) Spectroscopically resolved relative current vs. wavelength of a ZnO-DWCNT FET. The current starts to change at the absorption wavelength of ZnO. The wavelength was scanned with 3 s/nm at a bias of V$_{DS}$ = +1 V and the gate voltage was held at 0 V. The measurement was performed in vacuum. The inset shows the absorption spectrum of ZnO-DWCNT composites in methanol.}}
\end{figure}

\section*{Results and discussion}

First, we developed a synthetic route which avoids the use of stabilizing agents in order to exclude additional effects of the insulating organic molecules. The starting point for the synthesis of ZnO-DWCNT composites was a method introduced by Pacholski et al. \cite{9}, where spherical ZnO nanoparticles are obtained in methanol within a few minutes. After nucleation, nanorods form following a mechanism of oriented attachment. Long reaction times (up to 3 days) assure the merging of all nanoparticles into ZnO nanorod structures. It is also known that by decreasing the precursor concentration the formation of nanorods is suppressed and spherical particles are preserved \cite{9}.  In order to obtain ZnO-CNT composites we added DWCNTs as purchased (without further functionalization) to the zinc precursor prior to the hydrolysis by potassium hydroxide. To prevent nanotube bundling, which is often observed for small diameter tubes, methanol was substituted by 2-phenyl ethanol, a better dispersant medium for DWCNTs. Subsequently, the zinc precursor concentration had to be reduced due to its low solubility in 2-phenyl ethanol. This fact has two consequences: first, spherical ZnO particles are formed as mentioned before, and second, the reaction rate is increased due to lower zinc salt solubility \cite{52}. Since zinc containing material appears surrounding the carbon lattice immediately upon KOH injection, ZnO crystallites are assumed to nucleate directly on the nanotube. The reported adsorption of Zn$^{2+}$ ions on CNTs \cite{53,54} supports this assumption. Figures 1a and 1b show TEM images of DWCNTs decorated with ZnO nanoparticles with an average size of 5.3$\pm$0.8 nm. In the high resolution TEM image (Figure 1b), the two walls of the carbon nanotube as well as ZnO crystal planes are clearly observable. The crystallinity of the ZnO particles was additionally confirmed by X-ray diffractometry. Figure 1c shows an SEM image of a single DWCNT decorated with ZnO nanoparticles in between two gold leads defined by electron-beam lithography forming a FET device.

Raman spectroscopy provides information about the ratio of sp$^{3}$ to sp$^{2}$ hybridized carbon in CNTs. Figure 2a shows the Raman spectra of both raw DWCNTs before the decoration with ZnO nanoparticles and after the synthetic process. As can be observed, the samples synthesized at 60 $^{\circ}$C using 2-phenyl ethanol showed similar ratio between the weak D peak (which stands for the relative amount of sp$^{3}$ hybridized carbon) at 1363 1/cm  and the G peak (which represents the sp$^{2}$ hybridized carbon) at 1585 1/cm \cite{55,56}. This let us assume that there is no substantial covalent functionalization of the pristine DWCNTs and upon ZnO attachment.

The optical absorption edge of the nanoparticle-nanotube composites shown in the inset of Figure 2b lies at about 350 nm, which is somewhat smaller than the bulk value of ZnO. Taking into account that the Bohr exciton radius for ZnO lies around 2 nm \cite{57}, weak quantum confinement is observed for 5-6 nm ZnO nanoparticles. The nanoparticle size determined from the absorption spectrum \cite{58} matches well with that observed by transmission electron microscopy.

The composites of ZnO nanoparticles attached to DWCNTs were also investigated in terms of their electrical transport properties. Therefore, several FET devices were built by contacting either an individual DWCNT (blank device) or an individual ZnO-DWCNT composite with gold leads. Roughly one third of DWCNTs possess metallic properties. Thus, we chose those devices for our investigations which showed clear semiconductor behaviour. In the case of the composite material the ZnO-NPs were removed at the contact area to ensure intimate contact between DWCNT and lead metal (for details see the methods section above). Without removal of ZnO-NPs at the contacts we measured a maximum conductivity of 10$^{-11}$ A (without illumination), which is five orders of magnitude lower than the one of DWCNT or ZnO-DWCNT with ZnO removed at the contacts under the same conditions. Thus, we can exclude significant charge transport through the ZnO-NP layer. 

The work function of the here used DWCNTs varies in a small range around 4.9 eV \cite{59}. Since the gold leads exhibit a work function of 5.3 eV which is higher than the one of the CNTs a Schottky barrier is formed at the interface between the semiconducting nanotubes and the gold contacts. Due to the pinning in air \cite{60}, this leads to hole conduction and the device shows p-type characteristics. Both, blank devices containing only DWCNT as well as composites (Figure 3a) show hole conductivity without illumination. The appearing hysteresis in transfer characteristics is attributed to water adsorption and organic contamination from lithography process according to the literature \cite{61,62,63,64}. Consequently, reduced hysteresis under low pressure condition (10$^{-5}$ mbar) can be explained in terms of water desorption \cite{61}. The ZnO-DWCNT devices exhibit reduced on/off ratio compared to the blank devices, which can be understood in terms of screening of the tube by the charges of the n-type ZnO. Since the DWCNT channel is uniformly covered by ZnO nanocrystals, an additional layer is present between the Si/SiO$_{2}$ substrate and the gate, which lowers the efficiency of the applied gate voltage.

\begin{figure}[ht]
  \centering
  \includegraphics[width=0.45\textwidth]{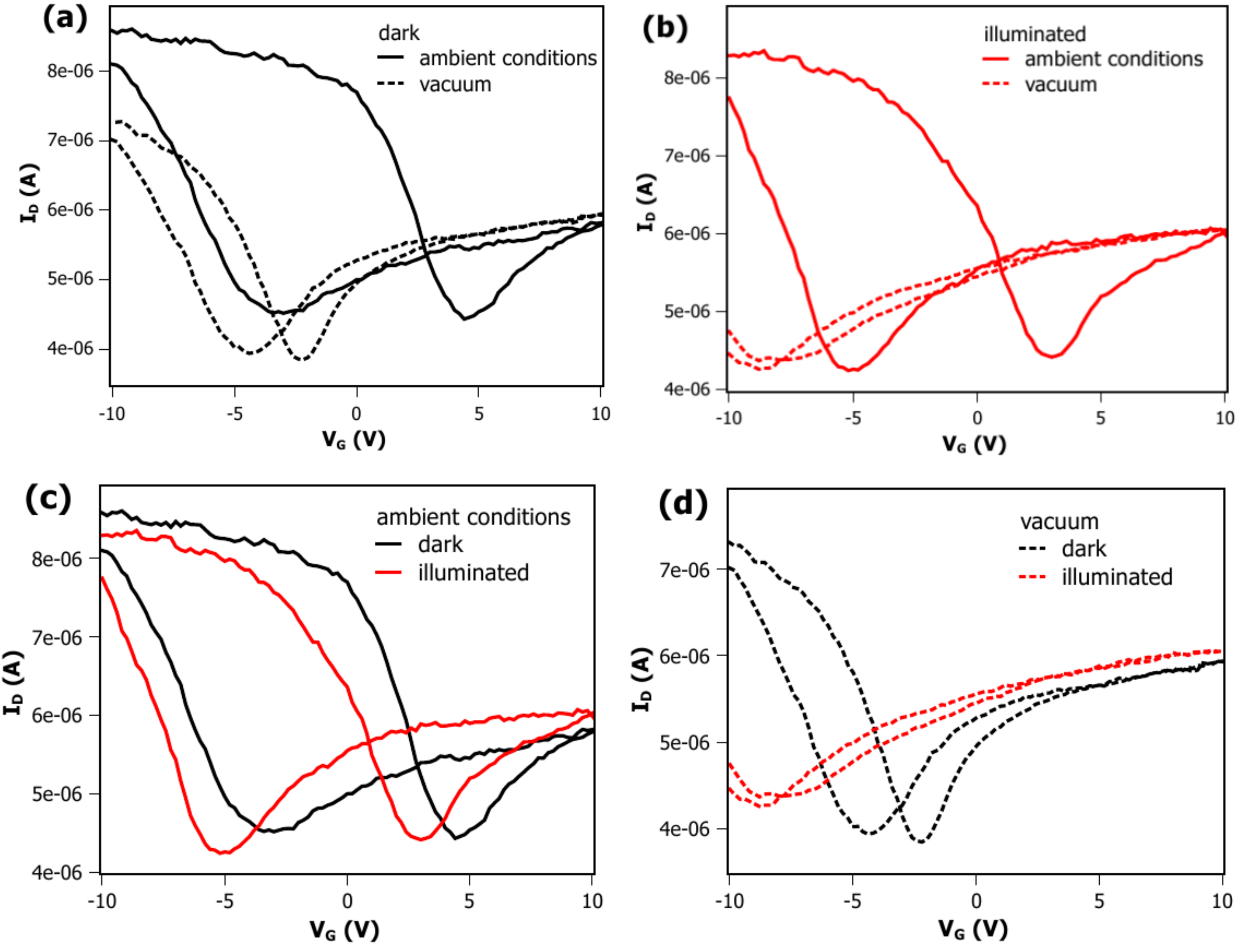}
  \caption{\textit{Comparison of ZnO-DWCNT-FET transfer characteristics under ambient condition (full line) and under low pressure condition (dotted line) (a) without and (b) upon white light illumination (Xenon lamp, 400 W). The threshold voltage shift after change of conditions is much larger with illumination (b) than without (a). Comparison of ZnO-DWCNT-FET transfer characteristics  with (red line) and without illumination (black line) (c) under ambient and (d) reduced pressure conditions shows that the threshold voltage shift upon illumination is much larger under vacuum (d) due to the lack of oxygen. All measurements were performed at a drain-source bias of VDS = +1 V.}}
\end{figure}

Due to strong oxygen affinity ZnO adsorbs O$_{2}$ readily forming O$_{2}^{-}$ species at the ZnO surface. Thus, oxygen is bound to the n-type ZnO by accepting the negative charge. In addition to the negative charge transfer from nanoparticles to the oxygen molecules, the electron transfer to the CNTs in ZnO-DWCNT composites is possible, since the Fermi level in n-doped ZnO is close to its conduction band level which lies at about -4.5 eV with respect to the vacuum level \cite{65} and thus above the Fermi level of the nanotubes. In vacuum, we expect the charge transfer from n-type ZnO to CNTs to be favoured due to oxygen absence. The comparison of recorded transfer characteristics of ZnO-DWCNT devices without light in Figure 3a displays a threshold voltage shift of more than 1 V to the negative by switching from ambient to low pressure condition, indicating the negative charge transfer from ZnO to CNTs and thus n-doping of the FET channel, as demonstrated by Heinze et al. for potassium \cite{66}. That means that by oxygen exclusion the charge transfer from ZnO to CNT becomes visible in transfer characteristics and the ZnO-DWCNT FETs are sensitive to oxygen. Comparable behaviour was not observed for pure DWCNT FETs.

In Figure 3b, where the transfer characteristics of illuminated devices under different conditions are displayed, an enhancement of the doping effect is indicated by a stronger threshold voltage shift. Under illumination, photo-generated electrons in ZnO are shown to follow three competitive paths in the presence of adsorbed oxygen \cite{67}. The electron can either recombine with the hole or with deep trapped holes or it can migrate to surface states. Furthermore, as described earlier, CNTs are able to accept electrons from ZnO due to favourable band alignment in the composite material. Thus, photo-generated electrons can easily pass from the nanoparticles to the closely attached nanotubes, as it could be shown for other NP-CNT systems \cite{34,38,39,40,41,42,45}. The hole might remain on the NP because of an unfavourable alignment of the hole state in the ZnO compared to the Fermi level of CNTs. Additionally, the n-type character of the ZnO stabilizes the positive charge on the NP.

\begin{figure}[ht]
  \centering
  \includegraphics[width=0.45\textwidth]{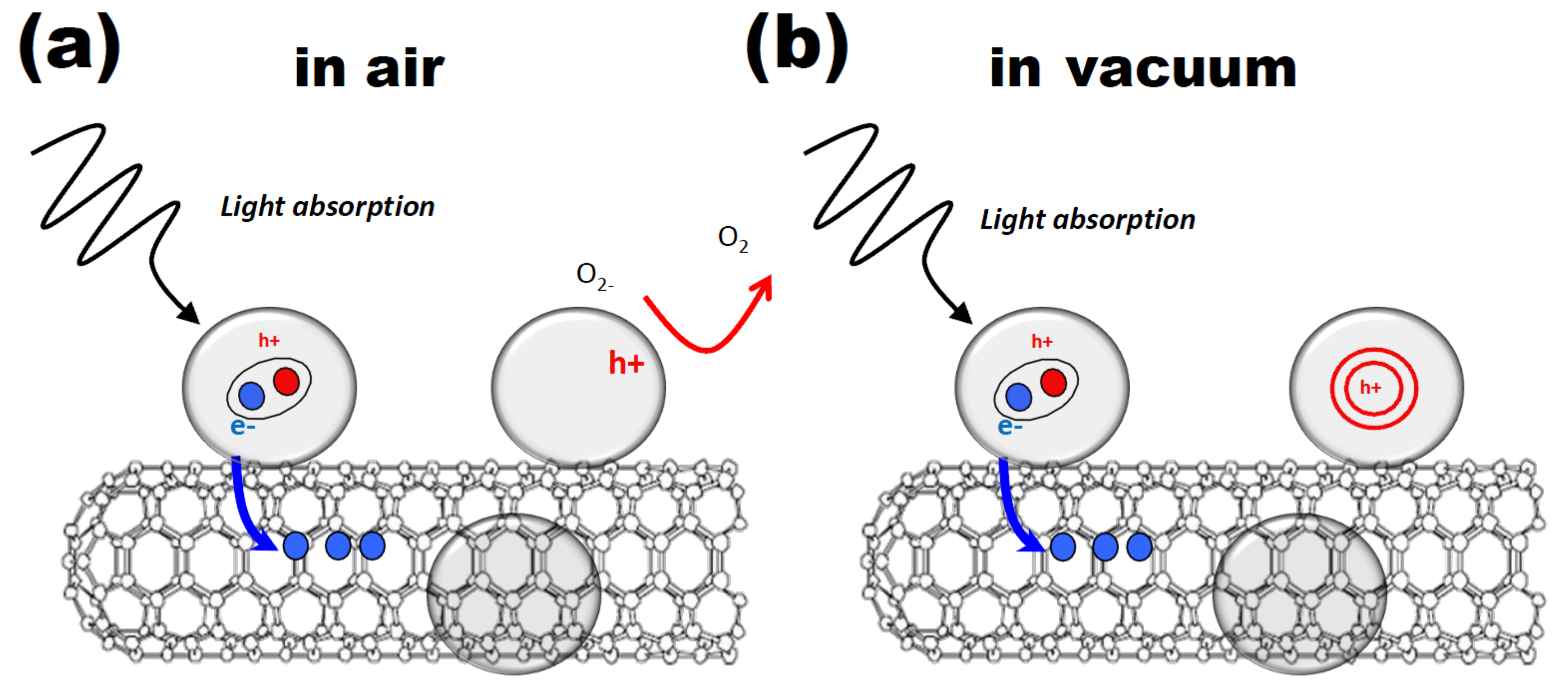}
  \caption{\textit{Proposed mechanism for the photo-gating effect for ZnO-DWCNTs (a) under ambient and (b) reduced pressure conditions. For simplicity, a single-wall CNT instead of a double-wall CNT is depicted.}}
\end{figure}

Under ambient conditions a small threshold voltage shift of about 1 V toward negative gate voltages as well as an increase in current at 0 V is observed upon illumination (see Figure 3c) indicating weak photo-n-doping \cite{66}. By switching to low pressure condition and exposing the same device to light the threshold voltage shift increases by a factor of three (see Figure 3d). While nanotube photo-doping is competitive with the processes described above under ambient conditions (Figure 3c), it seems to be the main path under reduced pressure as the transfer characteristics indicate in Figure 3d.

The fact that the strong improvement of carbon nanotube n-doping upon illumination happens by switching from ambient to low pressure conditions is understood in terms of oxygen desorption from attached ZnO (Figure 4). As mentioned above, oxygen is known to easily chemisorb on the ZnO surface. In the dark oxygen molecules can accept an electron from the n-type ZnO forming the O$_{2}^{-}$ specie \cite{68}. After the electron-hole pairs are generated by photon absorption, the holes migrate to the surface and discharge the O$_{2}^{-}$ species, while the remaining electrons can either recombine with the remaining holes or migrate to the surface where they assist to the re-adsorption of oxygen. Additionally, in the case of ZnO-CNT composite negative charge can be transferred to nanotubes. Consequently, when no oxygen is present the electron can either recombine with the hole or it can contribute to the CNT n-doping, while the hole remains stable in the particle since there are no O$_{2}^{-}$ species available. The positive charged particles can be considered as an additional source of gate voltage which is very efficient since the charge is located directly at the tubes. 

Further evidence for the photo-gating effect of ZnO on DWCNTs is found in the spectroscopically resolved photoconductivity measurements (Figure 2b), where a corresponding change in current occurs at wavelengths shorter than 350 nm, the absorption edge of ZnO (Figure 2b inset). In vacuum, the current sharply increases by up to 15 \% meaning, that the device is reasonably sensitive to UV-light.

\section*{Conclusions}

ZnO nanoparticles were grown directly on the non-functionalized carbon lattice of DWCNTs excluding long chain ligands at the NP-CNT interface by introducing the CNTs into a colloidal synthetic procedure. This methodology allows the formation of uniform and high quality NP-CNT composites with high coverage. The composites have been morphologically characterized and their gate response has been addressed and discussed in terms of a doping effect caused by closely attached n-type material and doping enhancement by illumination. Thus, ZnO-DWCNT FETs were shown to be oxygen and light sensitive devices demonstrating the potential of nano-composites for sensing and optoelectronic applications.


\end{document}